\documentclass[runningheads]{llncs}
\usepackage[T1]{fontenc}
\usepackage{graphicx}
\usepackage{graphicx}
\usepackage{colortbl}
\usepackage{booktabs}
\usepackage{color}
\usepackage{cite}
\usepackage{amsmath,amssymb,amsfonts}
\usepackage{graphicx}
\usepackage{textcomp}
\usepackage{xcolor}
\usepackage{url}
\usepackage{multirow}
\usepackage{times}
\usepackage{latexsym}
\usepackage{todonotes}
\usepackage{multirow}
\usepackage{graphicx}
\usepackage{amsmath}
\usepackage{amsfonts}
\usepackage{amssymb}
\usepackage{caption}
\usepackage[T1]{fontenc}
\usepackage{pifont}
\usepackage{subcaption}
\usepackage{booktabs}
\usepackage{colortbl}
\usepackage{amsmath} 
\usepackage{tikz}
\usepackage{xcolor}

\definecolor{goldcolor}{RGB}{184,134,11}
\definecolor{lightbeige}{RGB}{245,245,220}
\definecolor{lightgray}{RGB}{211,211,211}
\definecolor{lightmint}{RGB}{176,224,230}
\definecolor{darkmint}{RGB}{96,168,154}

\usepackage{enumitem}
\usepackage{amssymb}
\usepackage{pifont}
\usepackage{graphicx}

\usepackage[utf8]{inputenc}

\usepackage{microtype}

\usepackage{inconsolata}
\definecolor{LightCyan}{rgb}{0.88,1,1}
\usepackage{graphicx}
\usepackage{subcaption}
\usepackage{enumitem}
\usepackage{xcolor}
\usepackage{color, colortbl}
\usepackage[utf8]{inputenc}
\usepackage{times}
\usepackage{latexsym}
\usepackage{comment}
\usepackage[utf8]{inputenc}
\usepackage{subcaption}
\usepackage[T1]{fontenc}
\usepackage{microtype}
\usepackage{algpseudocode}
\usepackage{times}
\usepackage{latexsym}
\usepackage{amsmath}
\usepackage{pgfplots}
\usepackage{svg}
\usepackage{appendix}
\usepackage{makecell}
\usepackage{amsmath}
\usepackage{tikz}
\usetikzlibrary{positioning}
\usetikzlibrary{backgrounds}
\usepackage{tikzscale}
\usepackage{pgfplots}
\pgfplotsset{width=10cm,compat=1.9}
\usepgfplotslibrary{groupplots}
\usetikzlibrary{pgfplots.statistics}
\usetikzlibrary{patterns}
\usepackage{subcaption}
\usepackage{multirow}
\usepackage{adjustbox}
\usepackage{enumitem}
\usepackage{comment}
\usepgfplotslibrary{groupplots}
\usepackage[T1]{fontenc}
\definecolor{g1}{rgb}{0,0.8,0.4}
\definecolor{g4}{rgb}{0.88,1,0.88}
\definecolor{g2}{rgb}{0.66,1,0.66}
\definecolor{g3}{rgb}{0.8,1,0.8}

\definecolor{r1}{rgb}{1.0, 0.03, 0.0}
\colorlet{r2}{r1!50}
\colorlet{r3}{r1!30}
\colorlet{r4}{r1!15}
\def\BibTeX{{\rm B\kern-.05em{\sc i\kern-.025em b}\kern-.08em
    T\kern-.1667em\lower.7ex\hbox{E}\kern-.125emX}}

\begin{document}
\title{\texttt{PlotEdit}: Natural Language-Driven Accessible Chart Editing in PDFs via Multimodal LLM Agents}
%
%

\titlerunning{\texttt{PlotEdit}}

\author{Kanika Goswami\inst{1} \and
Puneet Mathur\inst{2} \and
Ryan Rossi \inst{2} \and
Franck Dernoncourt \inst{2}}
\institute{IGDTUW, India\textsuperscript{1} \email{goswami.kanika.96@gmail.com} \and Adobe Research\textsuperscript{2}}
\authorrunning{Goswami K. et al.}
%
%
\maketitle              

\begin{abstract}
Chart visualizations, while essential for data interpretation and communication, are predominantly accessible only as images in PDFs, lacking source data tables and stylistic information. To enable effective editing of charts in PDFs or digital scans, we present \texttt{PlotEdit} - a novel multi-agent framework for natural language-driven end-to-end chart image editing via self-reflective LLM agents. \texttt{PlotEdit} orchestrates five LLM agents: (1) Chart2Table for data table extraction, (2) Chart2Vision for style attribute identification, (3) Chart2Code for retrieving rendering code, (4) Instruction Decomposition Agent for parsing user requests into executable steps, and (5) Multimodal Editing Agent for implementing nuanced chart component modifications—all coordinated through multimodal feedback to maintain visual fidelity. \texttt{PlotEdit} outperforms existing baselines on the ChartCraft dataset across style, layout, format, and data-centric edits, enhancing accessibility for visually challenged users and improving novice productivity.

\end{abstract}

\section{Introduction}

\begin{figure*}
\centering
\scalebox{1}{
\includegraphics[width=1\textwidth]{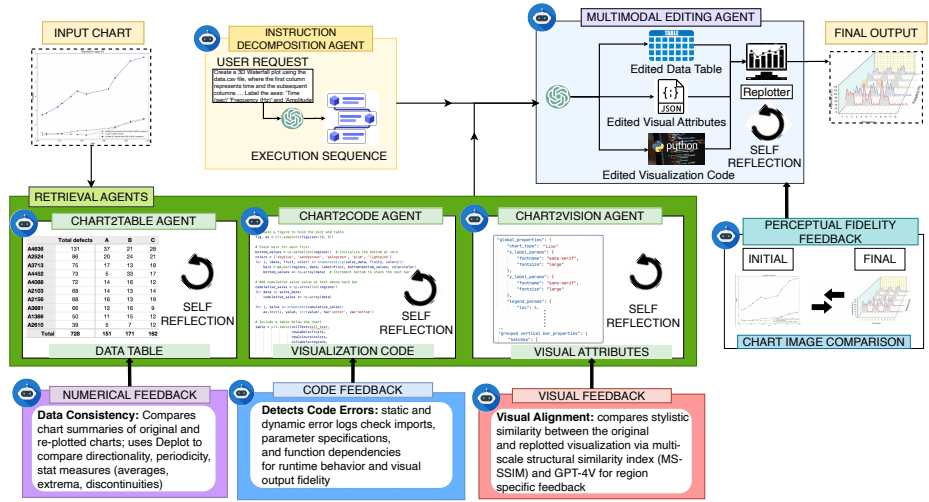}}
\caption{\small{\texttt{PlotEdit} accurately edits chart images as per user requests by orchestrating LLM agents: (1) Chart2Table for data table extraction, (2) Chart2Vision for style attribute identification, (3) Chart2Code for retrieving rendering code, (4) Instruction Decomposition Agent for parsing user requests into executable steps, and (5) Multimodal Editing Agent for implementing nuanced chart component modifications coordinated through multimodal feedback to maintain visual fidelity.}}
\label{fig:main}
\end{figure*}

Charts are designed with specific aesthetics and formats to effectively visualize tabular data, enabling the discovery of meaningful insights and the presentation of evidence-based narratives\cite{Srinivasan2018OrkoFM}. Charts from PDFs or digital scans often require modification to enhance accessibility, including reconstruction of low-resolution scans, reformatting for publication compatibility, improving data readability, enriching chart content, and adjusting visual attributes like color schemes and contrast ratios for visually impaired users. This necessitates the development of chart editing methods to enable filtering specific data segments, data revision, format conversions, and style alterations \cite{Narechania2020NL4DVAT}.

Traditional chart-editing methods present significant challenges, requiring extensive manual intervention, specialized software expertise, and access to source data which may be lost or unavailable \cite{Satyanarayan2018VegaLiteAG,Shao2020ChartDialogsPF}. While end-to-end vision-language models prove inadequate for complex 2D and 3D plots \cite{yan2024chartreformer}, direct LLM prompting often fails due to hallucinations in the chart de-rendering \cite{Han2023ChartLlamaAM}. To enhance the robustness and reliability of chart editing through language instructions, we investigate LLM-based agents that can interpret user intent, iteratively refine chart de-rendering, and implement precise adjustments.

To address these challenges, we propose \texttt{PlotEdit} (see Fig. \ref{fig:main}), a novel multi-agent framework designed to accurately edit chart images based on user specifications by utilizing five LLM agents. The framework comprises of three retrieval agents: (1) Chart2Table extracts data tables, (2) Chart2Vision identifies stylistic attributes, and (3) Chart2Code retrieves rendering code. \texttt{PlotEdit} employs three unique multimodal feedback signals—numeric, visual, and code—to rectify chart de-rendering errors through self-reflection. (4) Instruction Decomposition Agent parses complex user requests into executable steps, while a (5) Multimodal Editing Agent implements fine-grained modifications to the retrieved components. The editing agent is guided by perceptual fidelity feedback to ensure alignment with user specifications while maintaining the visual integrity of the original chart. \texttt{PlotEdit} significantly outperforms strong multimodal Transformer and LLM inference baselines by 9-14\% on the ChartCraft dataset across varied edit types, improving accessibility for visually challenged users, democratizing chart editing for novice users, and enhancing professional productivity.

\section{Methodology}

\noindent\textbf{Task Description}: Given an input chart figure $f$ and a user-specified editing request $r$ to modify plot attributes (including style, format, layout, or underlying data), the chart image editing task aims to generate output figure $\hat{f}$ that accurately implements the user's intended modifications while preserving the integrity of the original plot. To address this complex task, we propose \texttt{PlotEdit} that leverages a framework of five LLM agents:

\noindent\textbf{Retrieval Agents}: Chart figures predominantly exist in scanned images or digitized PDF files, making precise chart de-rendering a critical prerequisite for effective chart image editing. Our approach leverages GPT-4V to perform structural chart decomposition, simultaneously extracting three key components: underlying data table, visual attributes, and corresponding Python visualization code. The framework implements three modality-specific retrieval agents, defined as follows: \textbf{(1) Chart2Table Agent} leverages vision-language LLM capabilities to extract the data table through chain-of-thought prompting to predict count of rows/columns, headers fields, and cell values; \textbf{(2) Chart2Vision Agent} processes visual attributes into a JSON format, capturing fundamental features such as plot colors, styles (line, marker, bar patterns), font properties, titles, legend mappings; \textbf{(3) Chart2Code Agent} generates executable Python visualization code that specifies the chart type and layout components (axes grids, legends, and text labels).

\noindent\textbf{Self-Reflection via Multimodal Feedback Agents}: Data analysts frequently cross-reference visualization interfaces when interpreting charts through sensory feedback. We hypothesize that chart de-rendering should learn from mistakes and iteratively improve the retrieval of data tables, visual attributes, and rendering code. This necessitates that retrieval agents incorporate external feedback to rectify errors during de-rendering. While certain errors may be difficult to diagnose through textual analysis alone, they become apparent through visual comparison of original and replotted charts. To this end, we introduce three specialized feedback agents that enable iterative refinement:

\noindent\textbf{(i) Code Feedback} employs a dual-phase validation approach that provides structured feedback to the Chart2Code Agent for generating error-free code via self-reflection. The static validation phase leverages abstract syntax tree (AST) parsing to verify structural correctness and identify potential issues in library usage, parameter specifications, and function dependencies. The dynamic validation phase executes the code in a controlled environment to validate runtime behavior and visual output fidelity. Given a code snippet, the structured error log captures runtime error messages and warnings based on execution insights to ensure high quality of generated visualization code.

\noindent\textbf{(ii) Visual Feedback} compares stylistic attributes between the original chart $f$ and replotted visualization $f^{'}$ to guide the Chart2Vision Agent in refining visual attribute extraction. It segments both images into regions of interest (ROIs) and computes local similarity via multi-scale structural similarity index (MS-SSIM) \cite{wang2003multiscale} to identify visual discrepancies. For each spatial region in chart image, a multimodal LLM (GPT-4V) generates region-specific textual feedback based on the SSIM score to elucidate on the type and description of visual attribute mismatches. The detailed feedback ensures minimal deviation in decoded visual aspects.

\noindent\textbf{(iii) Numeric Feedback} validates data consistency by comparing the difference between LLM-generated chart summaries between the original chart and the replotted visualization to describe plot trends (directionality, periodicity) and structural relationships across data series. Further, it uses DePlot \cite{Liu2022DePlotOV} to extract local quantitative measures (averages, extrema, discontinuities) from both charts. Significant discrepancies between the summaries and statistical trends are analyzed by a text LLM (GPT-4o) to provide a description about potential errors in data extraction. This targeted feedback is used by Chart2Table Agent to improve data table accuracy and ensure faithful representation of the original chart's quantitative information. All three agentic feedback mechanisms are orchestrated sequentially until satisfactory results or exhaustion of max trials.

\noindent\textbf{(4) Instruction Decomposition Agent} uses GPT-4o to breakdown the user editing request into a sequence of executable steps via chain-of-thought prompting \cite{wei2022chain}, where each step corresponds to a specific modification targeting distinct plot components.

\noindent\textbf{(5) Multimodal Editing Agent}: Given the de-rendered chart image components and decomposed editing instructions, the Multimodal Editing Agent uses LLMs like GPT-4o to faithfully modifying the data table, visual attributes, and code according to user specification. It uses few-shot in-context learning for (i) \textit{Data Editing}: manipulates the extracted data table using pandas operations for data-centric edits such as range/series filtering or addition of new data points, (ii) \textit{Style Editing}: update the JSON  representation of foundational visual elements to implement style changes, (iii) \textit{Code Editing}: refactors the Python visualization code for both format conversion and layout re-composition.
\noindent\textbf{Perceptual Fidelity Feedback}: It is important that LLM hallucinations in editing agent do not significantly alter aspects of original chart figure that need to be preserved. To this end, Multimodal Editing Agent uses perceptual fidelity feedback to compare the edited visualization $\hat{f}$ against the original chart $f$ using GPT-4V to validate that only user-specified modifications have been implemented while maintaining fidelity of unchanged components. If discrepancies are found in regions that should remain unchanged, the editing agent receives region-specific SSIM feedback to revise its modifications preserving the original chart's integrity while facilitating tailored adjustments. A re-plotter software utilizes the edited chart components to reconstruct the output chart image.

\section{\texttt{PlotEdit} Evaluation and Conclusion}

\begin{table*}[ht]
\centering
\scriptsize
\begin{tabular}{l|ccc|ccc|ccc|ccc|ccc}
\toprule
\bf Edit Type & \multicolumn{3}{c|}{ \bf Style} & \multicolumn{3}{c|}{\bf Layout} & \multicolumn{3}{c|}{\bf Format} & \multicolumn{3}{c|}{\bf Data-Centric} & \multicolumn{3}{c}{\bf Overall} \\
\hline
\bf Method & SSIM & VAES & RMS & SSIM & VAES & RMS & SSIM & VAES & RMS & SSIM & VAES & RMS & SSIM & VAES & RMS \\
\hline
ChartLlama & 73.1 & - & - & 64.8 & - & - & 64.3 & - & - & 67.7 & - & - & 67.5 & - & - \\
ChartReformer & 83.3 & 86.6 & 90.4 & 82.6 & 89.7 & 89.2 & 84.2 & 90.5 & 91.2 & 81.4 & 84.2 & 88.4 & 82.4 & 86.3 & 89.3 \\
In-context Learning & 86.0 & 88.3 & 93.3 & 84.5 & 92.1 & 91.1 & 89.0 & 92.9 & 93.2 & 84.2 & 85.0 & 89.4 & 86.8 & 88.4 & 92.4 \\ \hline
\rowcolor{g3} \texttt{PlotEdit} & 87.3 & 90.5 & 96.5 & 91.3 & 95.2 & 93.4 & 91.2 & 93.5 & 95.0 & 87.5 & 88.7 & 92.6 & 89.0 & 91.5 & 93.8 \\ \hline
\rowcolor{r3} \texttt{PlotEdit} w/o MFA & 86.3 & 88.7 & 93.5 & 85.8 & 92.4 & 90.1 & 90.4 & 91.4 & 92.4 & 85.0 & 85.4 & 89.1 & 87.2 & 88.8 & 91.5 \\
\bottomrule
\end{tabular}
\caption{\label{tab:main_results} Performance of \texttt{PlotEdit} compared with baselines on ChartCraft dataset.}
\end{table*}

Following \cite{yan2024chartreformer}, we evaluate edited charts using three metrics: Structural Similarity Index Measure (SSIM) \cite{wang2004image} for structural fidelity, Relative Mapping Similarity (RMS) \cite{Liu2022DePlotOV} for data table accuracy, and Visual Attribute Edit Score (VAES) for style precision. Table \ref{tab:main_results} demonstrates that PlotEdit outperforms all baselines across metrics and edit categories. Empirical results indicate that LLM in-context learning underperforms compared to our agentic approach, primarily due to failures in the de-rendering process, particularly evident in data-centric and layout edits. ChartReformer \cite{yan2024chartreformer} exhibits notable limitations with data-centric edits, where systematic interpretation of input prompts and precise data manipulation are imperative. The superior efficacy of \texttt{PlotEdit} compared to ChartLLaMA \cite{Han2023ChartLlamaAM} and ChartReformer stems from its effective use of multimodal feedback mechanisms, facilitating enhanced data extraction precision and visual attribute recognition, as corroborated by our ablation analyses. The practical applications of 
\texttt{PlotEdit} lies in its application to chart editing within PDFs and scanned documents. While the current framework effectively translates user intent into initial chart modification suggestions, its integration with PDF readers presents opportunities for enhancing accessibility while streamlining document editing workflows, providing empirical validation of its practical capabilities.




\bibliographystyle{splncs04.bst}
\bibliography{references}

\begin{thebibliography}{10}
\providecommand{\url}[1]{\texttt{#1}}
\providecommand{\urlprefix}{URL }
\providecommand{\doi}[1]{https://doi.org/#1}

\bibitem{Han2023ChartLlamaAM}
Han, Y., Zhang, C.X., Chen, X., Yang, X., Wang, Z., Yu, G., Fu, B., Zhang, H.: Chartllama: A multimodal llm for chart understanding and generation. ArXiv  \textbf{abs/2311.16483} (2023), \url{https://api.semanticscholar.org/CorpusID:265466206}

\bibitem{Liu2022DePlotOV}
Liu, F., Eisenschlos, J.M., Piccinno, F., Krichene, S., Pang, C., Lee, K., Joshi, M., Chen, W., Collier, N., Altun, Y.: Deplot: One-shot visual language reasoning by plot-to-table translation. ArXiv  \textbf{abs/2212.10505} (2022), \url{https://api.semanticscholar.org/CorpusID:254877346}

\bibitem{Narechania2020NL4DVAT}
Narechania, A., Srinivasan, A., Stasko, J.T.: Nl4dv: A toolkit for generating analytic specifications for data visualization from natural language queries. IEEE Transactions on Visualization and Computer Graphics  \textbf{27},  369--379 (2020), \url{https://api.semanticscholar.org/CorpusID:221292836}

\bibitem{Satyanarayan2018VegaLiteAG}
Satyanarayan, A., Moritz, D., Wongsuphasawat, K., Heer, J.: Vega-lite: A grammar of interactive graphics. IEEE Transactions on Visualization and Computer Graphics  \textbf{23},  341--350 (2018), \url{https://api.semanticscholar.org/CorpusID:206805969}

\bibitem{Shao2020ChartDialogsPF}
Shao, Y., Nakashole, N.: Chartdialogs: Plotting from natural language instructions. In: Annual Meeting of the Association for Computational Linguistics (2020), \url{https://api.semanticscholar.org/CorpusID:218611161}

\bibitem{Srinivasan2018OrkoFM}
Srinivasan, A., Stasko, J.T.: Orko: Facilitating multimodal interaction for visual exploration and analysis of networks. IEEE Transactions on Visualization and Computer Graphics  \textbf{24},  511--521 (2018), \url{https://api.semanticscholar.org/CorpusID:2244239}

\bibitem{wang2004image}
Wang, Z., Bovik, A.C., Sheikh, H.R., Simoncelli, E.P.: Image quality assessment: from error visibility to structural similarity. IEEE transactions on image processing  \textbf{13}(4),  600--612 (2004)

\bibitem{wang2003multiscale}
Wang, Z., Simoncelli, E.P., Bovik, A.C.: Multiscale structural similarity for image quality assessment. In: The Thrity-Seventh Asilomar Conference on Signals, Systems \& Computers, 2003. vol.~2, pp. 1398--1402. Ieee (2003)

\bibitem{wei2022chain}
Wei, J., Wang, X., Schuurmans, D., Bosma, M., Xia, F., Chi, E., Le, Q.V., Zhou, D., et~al.: Chain-of-thought prompting elicits reasoning in large language models. Advances in neural information processing systems  \textbf{35},  24824--24837 (2022)

\bibitem{yan2024chartreformer}
Yan, P., Bhosale, M., Lal, J., Adhikari, B., Doermann, D.: Chartreformer: Natural language-driven chart image editing. In: International Conference on Document Analysis and Recognition. pp. 453--469. Springer (2024)

\end{thebibliography}
\end{document}